\documentclass[preprint,pre,aps,amsfonts,amsmath,showpacs,superscriptaddress,nofootinbib, draft]{revtex4}

\usepackage{bm}
\usepackage{amsmath}
\usepackage{graphicx}
\usepackage{amssymb,epsfig}
\usepackage{subfigure}
\begin{document}
\title{Percolation in the Harmonic
  Crystal and Voter Model in Three Dimensions}
\author{Vesselin I. Marinov}
\email{marinov@physics.rutgers.edu}
\affiliation{Department of Physics,   Rutgers University, Piscataway, New Jersey 08854-8019.}
\author{Joel L. Lebowitz}
\email{lebowitz@math.rutgers.edu}
\affiliation{Department of Mathematics and Department of Physics, Rutgers University, Piscataway, New Jersey 08854-8019.}
\date{\today}
\begin{abstract}
We investigate the site percolation transition in two strongly correlated systems in three dimensions: the massless
harmonic crystal and the voter model. In the first case we start with a Gibbs measure for the potential,
$U=\frac{J}{2} \sum_{<x,y>} \left(\phi(x) - \phi(y)\right)^2$, $x,y \in \mathbb{Z}^3$, $J > 0$ 
and $\phi(x) \in \mathbb{R}$, a scalar height variable, and
define occupation variables $\rho_h(x) =1,(0)$ for $\phi(x) > h (<h)$. The probability $p$ of a site being occupied, is then a function of $h$.
In the voter model we consider the stationary measure, in which each site is either occupied or empty, with probability $p$. 
In both cases the truncated pair correlation of the occupation variables, $G(x-y)$,  decays asymptotically
like $|x-y|^{-1}$. Using some novel Monte Carlo simulation methods and finite size scaling we find accurate values of $p_c$ as well 
as the critical exponents for these systems. The latter are different from that of independent percolation in $d=3$, as expected
from the work of Weinrib and Halperin [WH] for the percolation transition of systems with $G(r) \sim r^{-a}$ [A. Weinrib and B. Halperin, Phys. Rev. B 27, 413 (1983)].
In particular the correlation length exponent $\nu$ is very close to the predicted value of $2$ supporting the conjecture by WH that
$\nu= \frac{2}{a}$ is exact.
\end{abstract}

\pacs{05.70.Jk,05.50.+q,64.60.Ak,64.60.Fr,87.53.Wz,89.75.Da}
\maketitle
\section{Introduction}
A translation invariant ergodic system of point particles on a lattice, say $\mathbb{Z}^d$, in which each site is occupied with 
probability $p$, $0 \leq p \leq 1$, is said to percolate when it contains an infinite cluster of occupied sites,  connected
by nearest neighbor bonds. This event satisfies the zero/one law, i.e. the probability that the system 
percolates is either zero or one ~\cite{IntroPerc,Percolation}. For the case in which the sites are independent the transition 
from the non-percolating state for $p<p_c$ and the percolating  one for $p>p_c$ is one of the simplest examples
of critical phenomena. The probability 
that a given site, say the origin, is connected to infinity, i.e. is part of the infinite cluster, 
is zero for $p<p_c$ and strictly positive for $p>p_c$~\cite{IntroPerc,Percolation}. Much is known rigorously 
and even more from computer simulations and renormalization group calculations, about the 
nature of the percolation transition in the independent case. In particular it is known rigorously 
that $p_c$ is strictly greater than zero and less than one for $d \geq 2$ 
with $p_c(d)$ a decreasing function of $d$, etc. We also know explicitly or have bounds for some of the 
various exponents associated with the divergence of different quantities, e.g. the mean finite cluster size, 
when $p \rightarrow p_c$. We even know exactly the scaling limit of the shape
of the critical cluster on the triangular lattice ~\cite{smirnov}.
\\
\indent
It is generally believed that the critical properties, e.g. exponents, for independent percolation, but not $p_c$, are
universal: they do not depend on the particular lattice but only on the
dimensionality of the problem. The exponents are also believed not to be changed when 
one considers systems for which the occupation probabilities  for different sites are
not independent, as long as the correlations between occupied sites decay rapidly, say
exponentially ~\cite{WeinribHalperin}.  
\\
\indent
Less is known about the percolation transition when there are long
range correlations between occupied sites, e.g. when the correlations decay as a power law. Such power law decays
occur in many physical systems and the nature of the percolation transition in such systems has come
up recently in the study of  two dimensional turbulence ~\cite{turbulence}, and
of porous media, such as gels ~\cite{pacman}. 
\\
\indent
In a seminal work Weinrib and Halperin  ~\cite{Weinrib,WeinribHalperin}
argued that the critical exponents of the percolation transition should depend only on  the decay of 
the pair correlation 
$G(r)$ in such systems. In particular for $G(r) \sim r^{-a}$ the transition should be in a universality class which depends only on
$a$ and $d$. Their analysis was based on considering the variance of the particle density in a region of volume $\xi^d$, where $\xi$ is the
percolation correlation length which diverges as $p \nearrow  p_c$. They found that if $a<d$  these correlations are relevant if $a \nu - 2 < 0$. 
Here $\nu$ is the critical exponent which describes the divergence of the percolation correlation length $\xi$, e.g. the average radius of gyration 
of the clusters in the independent percolation problem as $p \nearrow p_c$, i.e. $\xi(p) \sim (p_c - p)^{-\nu}$.  
Weinrib and Halperin argued that systems that satisfy the above criteria belong to a new universality class for which the percolation correlation 
length exponent is $\nu_{long} = \frac{2}{a}$ ~\cite{Weinrib,WeinribHalperin}. They also checked this using a renormalization
group double expansion in $\epsilon =6-d$ and in $\delta = 4-a$. While the computations of WH were done only in the one loop approximation the exponent 
$\nu_{long}$ was conjectured to be exact ~\cite{Weinrib,WeinribHalperin}. 
\\
\indent
As pointed out by WH their results are consistent with those based on renormalization group ideas, both in real and momentum space, on the percolation of 
like-pointing Ising type spins at the critical point, see ~\cite{Weinrib} and references in there. 
There have also been some numerical tests of the WH predictions. For $d=2$ Prakash et al.~\cite{EStanley} have carried out  
Monte Carlo simulations for  percolation on artificially generated power law correlated occupation probabilities 
on $\mathbb{Z}^2$. This study confirmed the predictions of Weinrib and Halperin.
The only direct check of the WH prediction in $d=3$ we are aware of is in ~\cite{pacman} where
the authors introduced a bond percolation model in $\mathbb{Z}^3$, called Pacman percolation. They argued that 
the pair correlation for their model decays as $r^{-a}$, with a close to 1,  and obtained critical 
exponents which are consistent with WH, but since $a$ was not known exactly the results are not fully conclusive. 
\\
\indent
In this paper we study the percolation transition  in three dimensions for two
systems in which  the long range correlations arise naturally from
the microscopic dynamics: the massless harmonic crystal and the voter model on
$\mathbb{Z}^3$. Both of these systems are known rigorously to have $G(r) \sim r^{-1}$. They also have other similarities but are intrisicaly quite
different. The existence and nature of the percolation transition in these systems is of interest in their own right. 
Using Monte Carlo simulations and finite size scaling we find the $p_c$ for both models. We also find that both models have the same critical
exponents as expected from the WH predictions of a long range percolation universality class. 
\\
\indent
For the massless harmonic crystal in $\mathbb{Z}^d$ we define site x to be occupied if the scalar displacement field  $\phi(x)$ 
is greater than some preassigned value $h$ and empty if $\phi(x) < h$. Percolation then corresponds to the existence of an
infinite level set contour for $\phi(x) < h$. The existence of percolation threshold, i.e. $0<p_c<1$,
 was proven by Bricmont, Lebowitz and Maes~\cite{LebowitzMaes} for $d=3$. There are however no previous calculations (known to us)
concerning the actual value of $p_c$ or of the critical exponents for this system. One expects intuitively that the $p_c$ will be smaller than the $p_c$ for 
independent percolation ,c.f. ~\cite{EStanley}, but we know of no proof for this. Similarly a proof that $p_c >0$ for the harmonic crystal in $d > 3$,
or for the an-harmonic crystal in $d \geq 3$ is still an open problem ~\cite{GG}. For $d \leq 2$, $\phi(x)$ is for any $h$, either 
plus or minus infinity, with probability 1, when the size of the system goes to infinity. Thus either all sites are occupied or all 
sites are empty.
\\
\indent
The voter model, often used for modeling various sociological and biological phenomena , is a lattice system in which a site
$x$ is occupied or empty according to whether the ``voter'' living there belongs to party A or B. Voters change their party
affiliations according to a well defined stochastic dynamics ~\cite{Ligget}. The stationary state of this model is not known explicitly
but many of its properties are known exactly. In particular it has many features in common with the harmonic crystal. Like the harmonic
crystal, the stationary state of the voter model is trivial in $d \leq 2$; all sites occupied or all sites empty. On the other 
hand any $p$ is possible on $\mathbb{Z}^d$ for $d \geq 3$, where the truncated pair correlation decays, as it does for the harmonic crystal, like $r^{-1}$. 
No proof of the existence of a $p_c > 0$ is known for this system, i.e. the system could in principle percolate for arbitrary small $p$. For examples
of systems where $p_c \leq \epsilon$ for any $\epsilon > 0 $ see ~\cite{Chayes}. 
\\
\indent
The outline of the rest of the paper is as follows. In Section 2 we present the simulation methods and results the massless harmonic crystal. 
In particular we find $p_c  = 0.16 \pm 0.01$.  
In section 3 we study the voter model. We present a new efficient algorithm for simulating this model and report the results from
its implementation. We find in particular that $p_c = 0.10 \pm 0.01$ compared with a $p_c \cong 0.16$ obtained in ~\cite{LebowitzSaleur} using
a less reliable method. We conclude the paper with a brief discussion of some open problems.
\section{The Harmonic Crystal}
\subsection{Formulation}
Let $x \in \mathbb{Z}^d$ designate the sites of a d-dimensional simple cubic lattice and $\phi(x)$ be the scalar 
displacement field at site $x$. The interaction potential in a box $\Lambda$ with specified boundary conditions 
(b.c.), e.g. $\phi(x) = 0$ for $x$ on the boundary of $\Lambda$, has the form
\begin{equation}
 U = \frac{1}{2} J \sum_{<x,y>}\left(\phi(x) - \phi(y)\right)^2 + \frac{1}{2} M^2 
 \sum \phi(x)^{2} \equiv \frac{1}{2} \sum_{x,y} \phi(x) C^{-1}(x,y) \phi(y) 
\end{equation}
where $J > 0$ and $M \geq 0$, $<x,y>$ indicates nearest neighbor pairs, $|x-y| = 1$, on $\mathbb{Z}^d$. 
The sum is over all sites in $\Lambda$ with the specified b.c. The Gibbs equilibrium distribution of the
$\{\phi(x)\}$  at a temperature $\beta^{-1}$, $\mu^{M}_{\Lambda}(\{\phi(x)\}) = Z^{-1}_{M,\Lambda} = e^{-\beta U}$ is then Gaussian with a covariance matrix $\beta\mathbf{C}$
which is well defined for $M > 0$.
\\ 
\indent
The infinite volume limit Gibbs measure $\mu^{M}$ obtained when $\Lambda \nearrow \mathbb{Z}^d$ is ,for $M > 0$ ,translation invariant, with $<\phi(x)> = 0$  
and is independent of the boundary conditions ~\cite{Gibbs}. When $M \rightarrow 0$ , $\mu^{M}$ does not 
exist for $d\leq 2$~\cite{Gibbs}. This is due to the fact that the fluctuations of the field, e.g  $<\phi(x)^2>$,  
become unbounded for these dimensions. However, for $d \geq 3$ the Gibbs measure $\mu$ obtained as the
limit of $\mu_{M}$ when $M \rightarrow 0$  is well defined. (It is the same as the infinite volume limit
of the measure in a box with $M=0$ and prescribed boundary values $\phi(x) = 0$). In this limit the pair 
correlations between different sites have the long distance behavior $\frac{1}{r^{d-2}}$, $r = |x-y|$ 
for $d>2$ ~\cite{Gibbs}. 
\\
\indent
Following \cite{LebowitzMaes} we define the occupation variable $\rho_h (x)$
\begin{equation}
\label{eqn:theta}
\rho_h(x)= 
\begin{cases}
1 & \text{if $\phi(x) \geq h$}\\
0 & \text{if $\phi(x) < h$}\\
\end{cases}
\end{equation}
and let $p=<\rho_h(x)>_{\mu^M}$, where the average is over the Gibbs measure $\mu^M$. We can also define a new 
measure $\hat{\mu}^M$ on the occupation variables $\rho_h(x)$={0,1} by a projection of $\mu^M$. All expectations 
involving a function of the occupation variables can be computed directly from $\hat{\mu}^M$. The correlations 
between the occupation variables have the same asymptotic decay properties as those of the field variables $\phi$,   
\begin{equation}
\label{rho}
\left<\rho_h(x)\rho_h(y)\right>_{\hat{\mu}^M} - p^2 \sim
\frac{e^{-|x-y|/{\xi_M}}}{|x-y|^{d-2}} \ \ \text{for $d>2$}
\end{equation} 
where $\xi_M \sim M^{-1}$ and the averages are with respect to $\hat{\mu}^M$(or $\mu^M$). In the limit 
$M \rightarrow 0$ the measure $\hat{\mu}$ has a pair correlation that decays like $r^{2-d}$ for $d > 2$. We 
note that $\hat{\mu}$ is not Gibbsian for any summable potential, c.f ~\cite{Sokal}. 
\subsection{Results}
\begin{figure}
\epsfxsize=4in
\begin{center}
\leavevmode
\epsffile{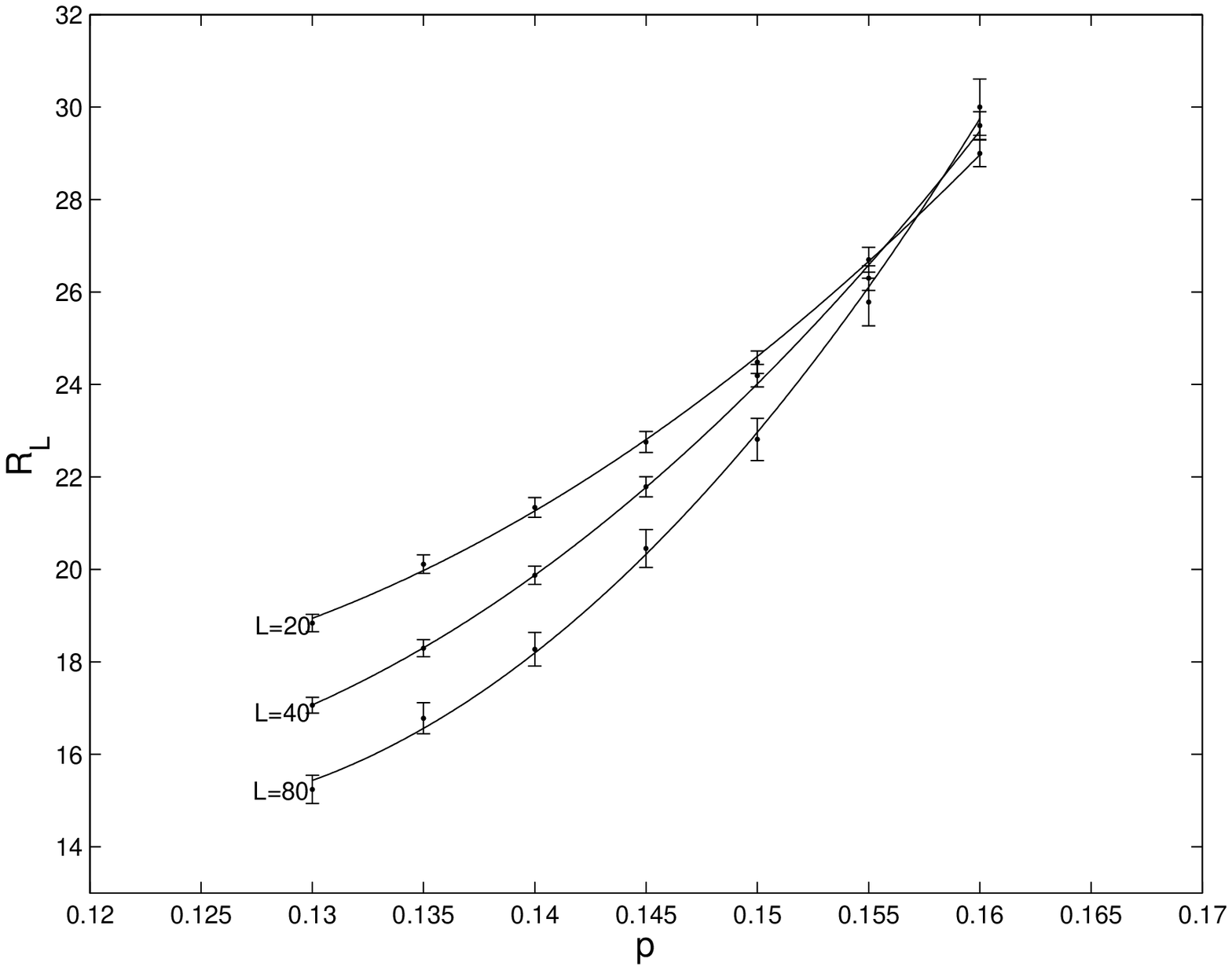}
\end{center}
\caption{Plot of $R_L$ versus p for the $d=3$ massless harmonic
  crystal. We estimate $p_c = 0.16 \pm 0.01$}
\label{fig:Fig1}
\end{figure}

\begin{figure}
\epsfxsize=4in
\begin{center}
\leavevmode
\epsffile{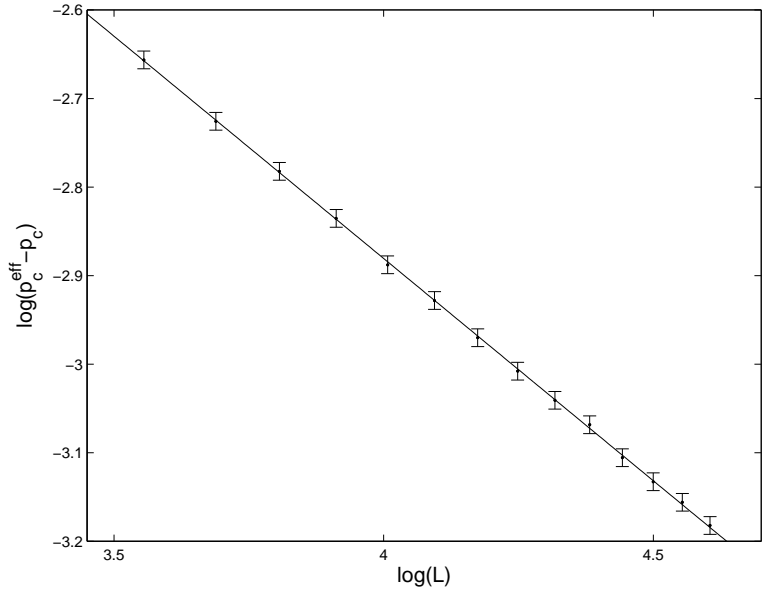}
\end{center}
\caption{Plot of $\log(p_c^{eff} - p_c)$ versus $\log(L)$. The slope
  if the straight line gives  $\nu = 2.00 \pm 0.04$ }
\label{fig:Fig2}
\end{figure}

\begin{figure}
\epsfxsize=4in
\begin{center}
\leavevmode
\epsffile{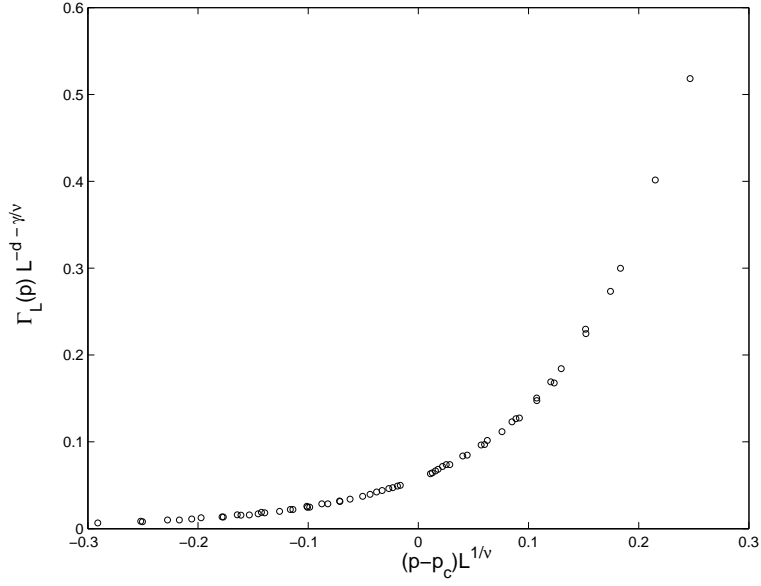}
\end{center}
\caption{Plot of $\Gamma_L L^{-d -\frac{\gamma}{\nu}}$ versus $(p-p_c)L^{\frac{1}{\nu}}$ for the $d=3$ massless
harmonic crystal for $p_c = 0.16$, $\nu = 2$ and $\frac{\gamma}{\nu}=1.8$. 
We have plotted data points for $L=30,60,120$ for $p=0.13$ to $0.16$ in steps of $0.005$ and for 
$L = 40,80,160$ for $p=0.13$ to $0.18$ in steps of $0.005$.}
\label{fig:Fig3}
\end{figure}

\begin{figure}
\epsfxsize=4in
\begin{center}
\leavevmode
\epsffile{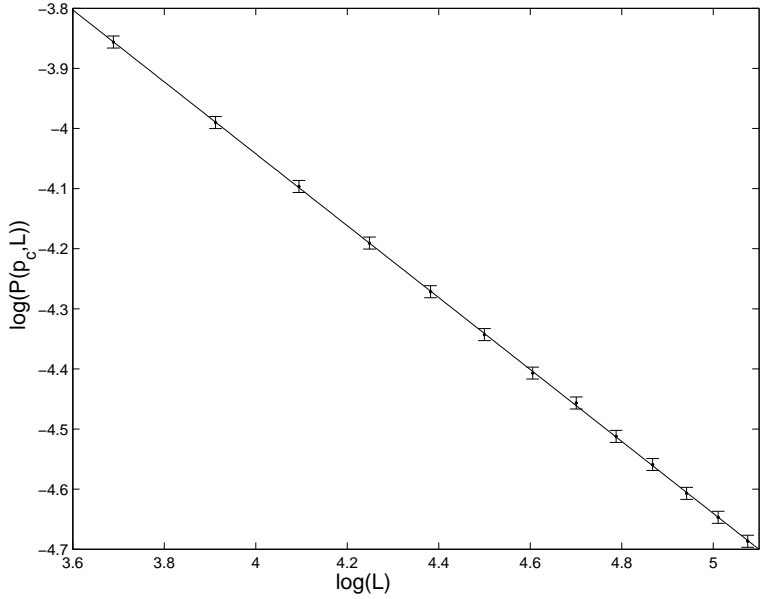}
\end{center}
\caption{Plot of $\log(P(p_c,L))$ versus $\log(L)$. The slope of the
  straight line gives $\frac{\beta}{\nu} = 0.60 \pm 0.01$}
\label{fig:Fig4}
\end{figure}

\indent
\\
Simulating the harmonic crystal on finite lattices is easy, the elements in a discrete Fourier transform of a 
harmonic crystal are independently distributed Gaussian random variables with easily computed variances~\cite{SSheffield}. 
We consider the system on a lattice with periodic boundary conditions and exclude the zero mode. This is essentially equivalent
to fixing $\left<\phi\right> = 0$.
\\
\indent
There are many methods for obtaining the percolation threshold using data obtained from simulations on 
finite systems~\cite{IntroPerc}. We used the method employed in
~\cite{MCTransf,LebowitzSaleur}. For a cube of linear size $L$ let 
\begin{equation}
\label{Gammadef}
\Gamma_L = \left<\sum_j j^2 n_j\right>
\end{equation}
where $n_j$ is the number of clusters of $j$ sites, defined by the occupation
variables $\rho_h(x)$, and the average is taken over a 
large number of samples obtained from simulation of the model. We
calculate  $\Gamma_L$ for different sizes $L$ and concentration 
of occupied sites $p$ defined as in (\ref{eqn:theta}). 
\\
\indent
One expects ~\cite{IntroPerc,MCTransf,Binder} that for large $L$,and
$(p_c-p) \ll 1$, $\Gamma_L$ should have a finite size scaling form, 
\begin{equation}
\label{Scaling}
L^{-d}\Gamma_L \sim L^{\frac{\gamma}{\nu}}F(L^{\frac{1}{\nu}}(p-p_c))  + \mbox{corrections to scaling}, 
\end{equation}
where $\gamma$ is the critical exponent for the divergence as $p \nearrow p_c$  of the
second moment of the cluster size distribution, defined as the limit $L \rightarrow \infty$ of $\frac{\Gamma_L}{L^d}$. 
Corrections to scaling should go to zero for $L \rightarrow \infty$.
\\
\indent
For $p > p_c$, for an infinite system the second moment of the cluster size distribution can be defined
by excluding the infinite cluster. This diverges with a critical exponent $\gamma^{'}$ for $p \searrow p_c$.
The finite system analog is $\frac{\Gamma_L^{'}}{L^d}$ which is defined similarly to $\frac{\Gamma_L}{L^d}$ 
but not including the spanning cluster. $\Gamma_L^{'}$ scales as 
\begin{equation}
\label{Scaling2}
L^{-d}\Gamma_L^{'} \sim L^{\frac{\gamma^{'}}{\nu}}F^{'}(L^{\frac{1}{\nu}}(p-p_c))  + \mbox{corrections to scaling}.
 \end{equation}
It is believed that $\gamma^{'}$ = $\gamma$.
\\
\indent
According to finite size scaling theory the number of sites in the largest cluster in a finite system of
linear size $L$, $P_L(p)$, scales for $|p-p_c| \ll 1$ as 
\begin{equation}
\label{beta}
P_L(p) \sim L^{d - \frac{\beta}{\nu}}G(L^{\frac{1}{\nu}}(p-p_c)) + \mbox{corrections to scaling}
\end{equation}
~\cite{IntroPerc,Binder}, where $\beta$ is the critical exponent for the approach to zero of the
fraction of sites belonging to the infinite cluster in an infinite
system as $p \searrow p_c$. Using the hyper-scaling relation $d = 2\frac{\beta}{\nu} + \frac{\gamma}{\nu}$ we see that 
(\ref{Scaling}), (\ref{Scaling2}) and (\ref{beta}) lead to the scaling form  (\ref{Scaling})
being valid for all $|p-p_c| \ll 1$ and large $L$. That is on a finite system we do not need to differentiate between 
$p<p_c$ or $p>p_c$, we may include all the clusters when calculating $\Gamma_L(p)$. 
\\
\indent
Assuming (\ref{Scaling}) is valid for $|p-p_c| \ll 1$  the ratio $R_L = \frac{\Gamma_{2L}}{\Gamma_L}$ should become
independent of $L$, for large $L$, at $p=p_c$. 
Plotting these ratios as a function of $p$ for different sizes $L$ and looking for the 
intersection of these different curves then yields $p_c$.  The value of the ratios at the intersection point of 
the $R_L$ curves should be equal to $2^{d + \frac{\gamma}{\nu}}$ giving us a way to measure $\frac{\gamma}{\nu}$.  
Moreover ,we also have  
\begin{equation}
\frac{1}{\nu} = \frac{\log{\left(\frac{dR_{2L}}{dp}/\frac{dR_L}{dp}\right)}}{\log {2}}.
\end{equation} 
Thus the  slopes of these curves should also give $\nu$. 
\\
\indent
In Fig.\,\ref{fig:Fig1} we present results of the simulation for the massless harmonic crystal on a cubic lattice 
with periodic boundary conditions. Each $\Gamma_L$ was averaged over
$48000$ samples except for $L=160$ where the average is over $2400$
samples. To determine the error bars we have divided the output of the
simulations into 10 parts and assuming that the averages are Gaussian
distributed we evaluated the variance which we used as a measure
of the uncertainty.
From the intersection of the curves, after interpolation, we obtain $p_c = 0.16 \pm 0.01$. 
Comparing the slopes of the $R_L$ curves for $L=80$ and $L=40$ we obtain $\nu = 2.1 \pm 0.5$.
From the value of $R_L$ at the intersection point of the curves we obtain $\frac{\gamma}{\nu} = 1.8 \pm 0.1$. 
We actually computed $\Gamma_L$ for the sequences $L=10,20,40,80,160$ and
$L=15,30,60,120$. All the simulation results are consistent with what is plotted in  Fig.\,\ref{fig:Fig1} where we have
used only part of these simulations since the plot is otherwise cluttered. These values clearly show that our system
is in a different universality class from independent percolation since for the latter 
$\nu   = 0.876 \pm 0.001$ and $\frac{\gamma}{\nu} = 2.045 \pm 0.001$ ~\cite{PercParisi}.
\\
\indent
The above method is good for finding the percolation threshold and the
ratio of critical exponents $\frac{\gamma}{\nu}$ but 
clearly does not give good results for $\nu$. To obtain more
precise result for the  percolation correlation length 
exponent we evaluated the probability that there is a ``wrapping
cluster'', i.e. one that wraps around the torus, 
for different densities $p$ of occupied sites and different linear sizes $L$.
\\
\indent
For fixed $L$ we denote by $p_c^{eff}$ the value of the density of occupied sites for which one half 
of the realizations will have such a wrapping cluster. This should obey the following scaling relation 
$p_c^{eff} - p_c \sim L^{-\frac{1}{\nu}}$~\cite{IntroPerc}. For sizes between $30$ and $100$  we evaluated
$p_c^{eff}$ from doing simulation in a range between $p=0.13$ and $p=0.25$ in steps of $0.005$. 
For each such system $24000$ samples were generated. The slope of
$\log(p_c^{eff} - p_c)$ versus $\log(L)$ should give us $\nu$. 
A plot of the results is presented in Fig.\,\ref{fig:Fig2}. The slope
of the fitted straight line is $0.50 \pm 0.01$ which gives $\nu = 2.00 \pm 0.04$.
This is in good agreement with the theoretical prediction $\nu = 2$ of
Weinrib and Halperin~\cite{WeinribHalperin}. 
\\
\indent
We have used the obtained values of $p_c$, $\frac{\gamma}{\nu}$ and $\nu$ to draw  Fig.\,\ref{fig:Fig3} where we 
see a good collapse of the data points to a smooth curve. 
\\
\indent
We also calculated the ratio of the critical exponents $\frac{\beta}{\nu}$. We did this by finding
the fraction of  sites  that belong to the largest cluster 
in a system of linear size $L$, $\frac{P(p_c,L)}{L^d}$,when we simulate at the
approximate critical density. From (\ref{beta}) we see that $\frac{P(p_c,L)}{L^d}
\sim L^{-\frac{\beta}{\nu}}$. The result for systems of size 
from $40$ to $170$ averaged  over $24000$ samples is presented in 
(Fig.\,\ref{fig:Fig4}). From the slope of the fitted  straight line we
obtain $\frac{\beta}{\nu} = 0.60 \pm 0.01$. Moreover, the fact that 
$P(p_c,L)$ follows well a power law behavior supports our 
contention that the true critical value is near $p_c = 0.16 \pm 0.01$. Observe also that  
$2\frac{\beta}{\nu} + \frac{\gamma}{\nu} = 3.0 \pm 0.2$ and thus the
hyper-scaling relation is satisfied.
\section{The Voter Model}
\subsection{Formulation}
\indent
Another system whose pair correlations decays like that of the massless harmonic crystal 
is the voter model in $\mathbb{Z}^d$ ~\cite{Ligget}. 
\\
\indent 
The voter model is defined through a stochastic time evolution.  Each
lattice site is occupied by a voter who can have two possible
opinions, say yes or no. With rate $\tau^{-1}$ the voter at site x adopts the opinion of one of his/her 
$2d$ neighbors chosen at random. More specifically letting $\rho(x) = 0,1$, 
$x \in \mathbb{Z}^d$, the time evolution of the voter model is
specified by giving the rate $C_v(x,\mathbf{\rho})$ for a change at 
site x when the configuration is given by $\mathbf{\rho}$
\[C_v(x, \mathbf{\rho}) = \frac{1}{\tau}\left[1 - \frac{1}{2d} (2\rho(x)-1) \sum_{|y-x|=1} (2\rho(y)-1)\right]\]
where $\tau$ sets the unit of time.
\\
\indent
It is clear that for the voter model on a finite set $\Lambda \subset
\mathbb{Z}^d$ with periodic or free boundary conditions (b.c.), there
will be only two possible stationary states: either $\rho(x) = 1$ 
or $\rho(x)=0$ for all $x \in \Lambda$. The same is true for the
voter model on an infinite lattice in one and two dimensions:
the only stationary states are the consensus states. However for $d
\geq 3$ there are, as for the massless harmonic crystal, 
unique stationary states for every density $p$ of positive spins, $p=\left<\rho(x)\right>$.
 The correlations in this state decay as
\[ \left<\rho(x) \rho(y)\right>-p^2 = p(1-p) G_d(x-y) \]
where $G_d(x)$ is the probability for a random walker, starting at $x
\in Z^d$ to hit the origin before escaping to infinity. 
It is well known that $G_d(x) \sim \frac{1}{|x|^{d-2}}$ for $d \geq 3$, i.e  the pair correlation for the
voter model has the same long range behavior as the massless harmonic crystal. 
\\
\indent
\subsection{Simulation Method}
An efficient method to simulate the voter model is to consider a box $B_{\mathcal{L}}$ of linear size ${\mathcal{L}}$ with stochastic boundary
conditions, i.e. when a voter looks at the boundary he sees $1$ with probability $p$ and $0$ with probability
$1-p$. It is then possible to show that the distribution of the configuration of voters in a box $B_L$ of 
size $L<\mathcal{L}$ centered inside $B_{\mathcal{L}}$ and far away from the boundary will approach the steady
state measure (restricted to $B_L$) with density p for the voter model when $\mathcal{L} \rightarrow \infty$. 
In order to sample from the measure for the voter model inside $B_{\mathcal{L}}$ with such stochastic 
boundary conditions we use the following algorithm: 
Start a random walk from each site of $B_L$ and let these random walks move independently until two of them
meet in which case they coalesce. When a random walk hits the boundary of $B_{\mathcal{L}}$ it is frozen. We continue this
until all the random walkers either coalesce or are frozen. After this is done we independently for each frozen walker,
assign the value $1$ with probability $p$ and the value $0$ with probability $1-p$, then assign that same value to its ancestors, that is all the random 
walkers that have coalesced with it. In this way we assign values 1 or zero to all the sites in $B_L$. 
One can prove that in this way we sample configurations inside $B_L$
with the distribution coming from the voter model in $B_{\cal{L}}$ with the stochastic boundary conditions
discussed above. The advantage of this way of simulating is that one is guaranteed that the sampling is from the 
steady state measure with these boundary conditions.
\subsection{Results}
\begin{figure}
\epsfxsize=4in
\begin{center}
\leavevmode
\epsffile{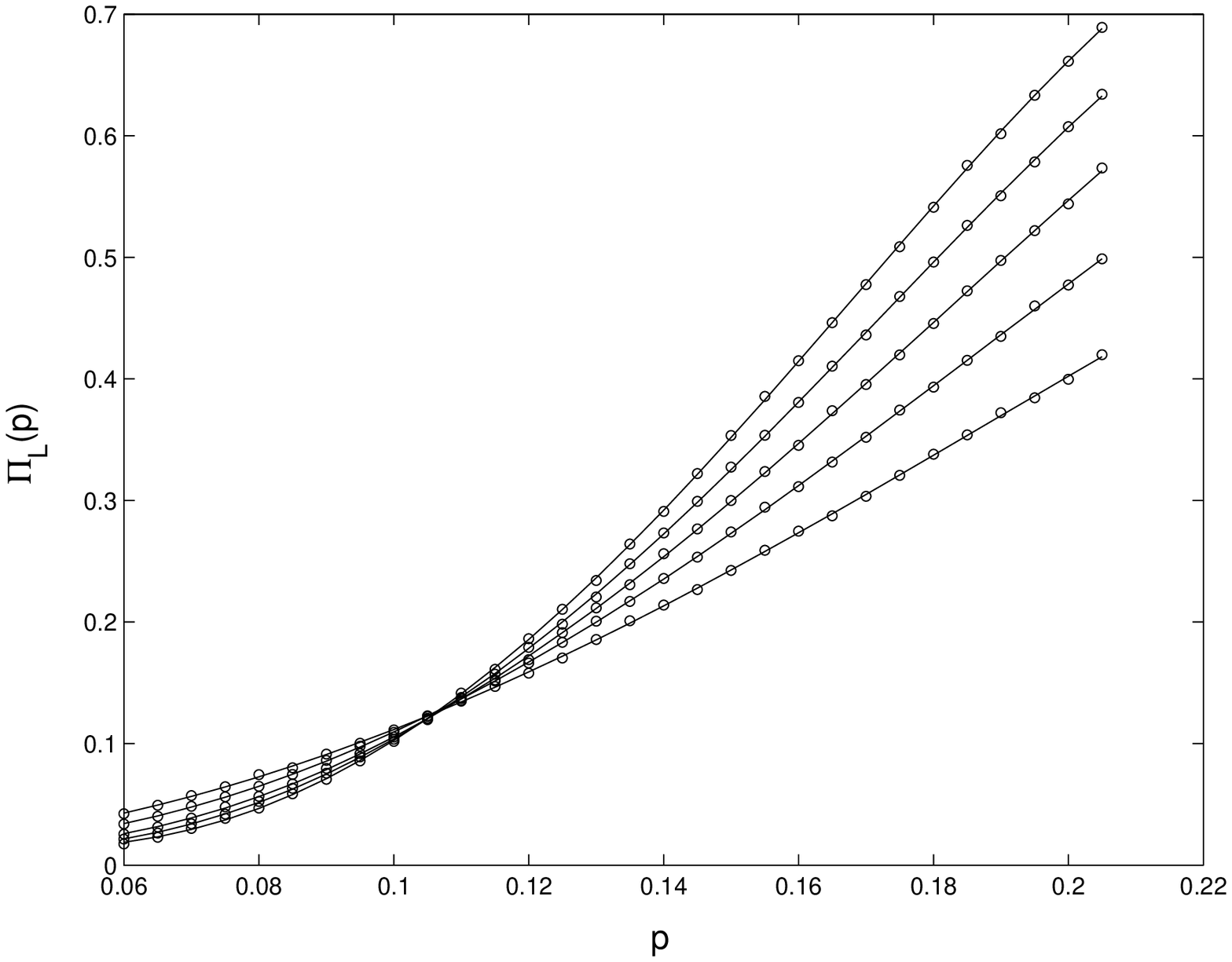}
\end{center}
\caption{Plot of $\Pi_L$ versus p for the $d=3$ voter model. We have plotted data points for $L=10,15,20,25$ and $30$
from $p=0.06$ to $p=0.205$ in steps of $0.005$. Each  point is an average over $10^5$ samples. The error bars are not 
shown since on this scale they are too small.}
\label{fig:Fig5}
\end{figure}

\begin{figure}
\epsfxsize=4in
\begin{center}
\leavevmode
\epsffile{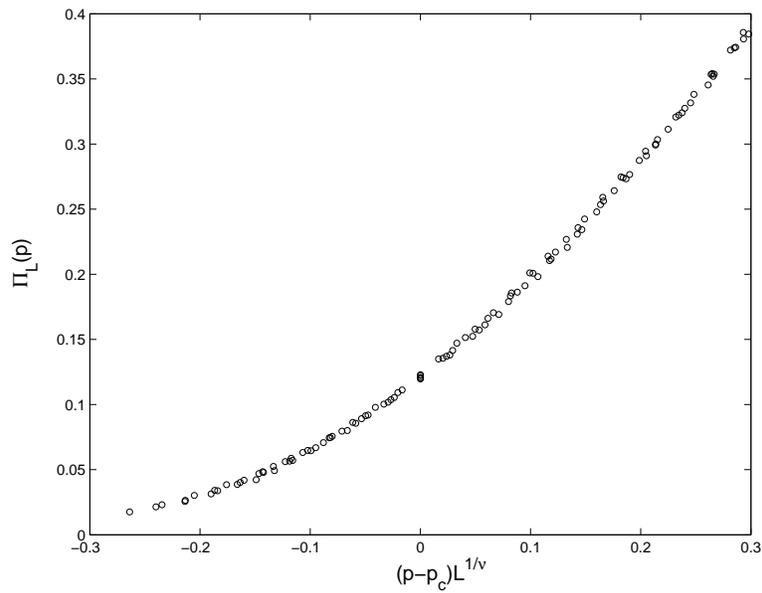}
\end{center}
\caption{Plot of $\Pi_L$ versus $(p-p_c)L^{\frac{1}{\nu}}$ for the $d=3$ voter model for $p_c = 0.105$ and $\nu = 2$.
We have used the same data that was used to create Fig.\,\ref{fig:Fig5}}
\label{fig:Fig6}
\end{figure}

\begin{figure}
\epsfxsize=4in
\begin{center}
\leavevmode
\epsffile{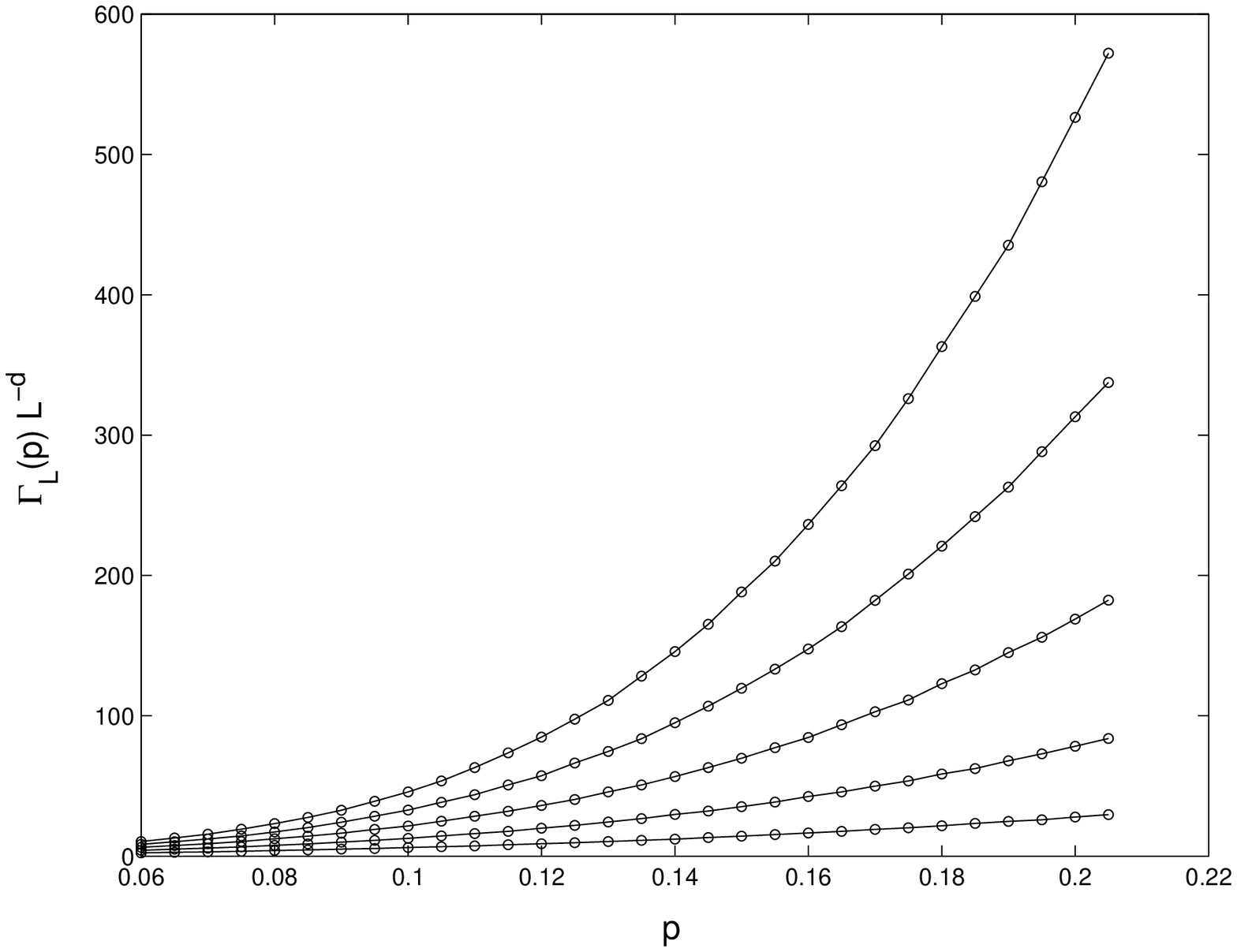}
\end{center}
\caption{Plot of $\Gamma_L L^{-d}$ versus $p$ for the $d=3$ voter model. We have plotted data points 
for $L=10,15,20,25$ and $30$ from $p=0.06$ to $p=0.205$ in steps of $0.005$. Each point is an average 
over $10^5$ samples. The error bars are not shown since on this scale they are too small.}
\label{fig:Fig7}
\end{figure}

\begin{figure}
\epsfxsize=4in
\begin{center}
\leavevmode
\epsffile{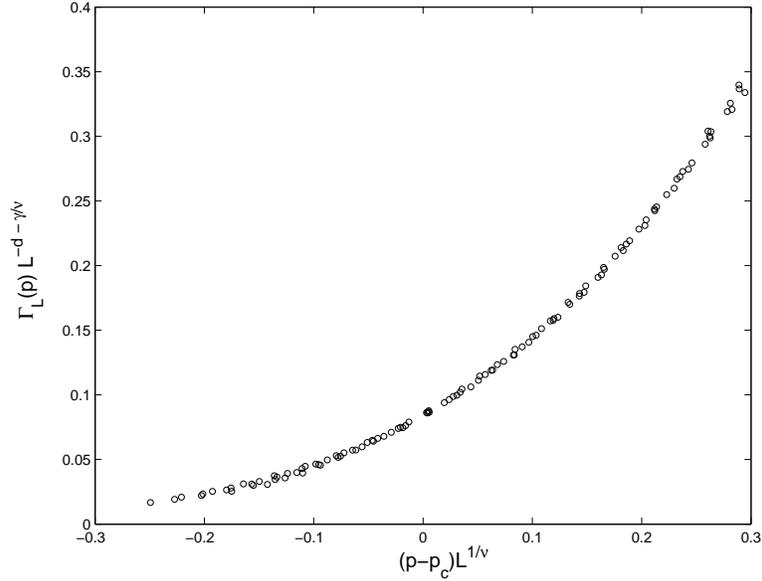}
\end{center}
\caption{Plot of $\Gamma_L L^{-d-\frac{\gamma}{\nu}}$ versus $(p-p_c)L^{\frac{1}{\nu}}$ for the $d=3$ voter model 
for $p_c = 0.105$, $\nu = 2$ and $\frac{\gamma}{\nu}=1.9$. We have used the same data that was used to create Fig.\,\ref{fig:Fig7}}
\label{fig:Fig8}
\end{figure}

Using this method of generating configurations inside $B_L$ for different
p we looked for a spanning cluster inside $B_L$. We did simulations for sizes $L=10,15,20,25$ and $30$ with $\mathcal{L}= 160$. 
The results which are the same for all $\mathcal{L}$ in the range $(120,160)$ are presented in Fig.\,\ref{fig:Fig5}. If we 
assume the scaling form for the spanning probability ~\cite{IntroPerc}
\begin{equation}
\Pi_L(p) = F((p-p_c)L^{1/\nu})
\end{equation}
then by collapsing the data Fig.\,\ref{fig:Fig6} we obtain $p_c = 0.10 \pm 0.01$ and $\nu = 2 \pm 0.2$.
\\
\indent
To find $\frac{\gamma}{\nu}$ we measured $\frac{\Gamma_L}{L^3}$ and we assume the scaling form (\ref{Scaling}). Note that 
in this case we do not have periodic boundary conditions. Results from the simulation are presented in Fig.\,\ref{fig:Fig7}.
Collapsing the data Fig.\,\ref{fig:Fig8} we obtain $p_c = 0.10 \pm 0.01$, $\frac{\gamma}{\nu} = 1.9 \pm 0.2$ and 
$\nu = 2 \pm 0.2$.
\\
\indent
Analogous simulation measurements for $P(p,L)$ gave $\frac{\beta}{\nu} = 0.6 \pm 0.1$. As in the case of the massless 
harmonic crystal the exponents we found satisfy the hyper-scaling relation  $2\frac{\beta}{\nu} + \frac{\gamma}{\nu}=d$.
The exponents for both the massless harmonic crystal and voter model seem to agree within the error bars. 
\subsection{Comparison of $p_c$ with previous simulations}
The percolation transition in the $d=3$ voter model was first investigated
in~\cite{LebowitzSaleur}. This was done by considering voters who
occasionally change their opinions spontaneously, i.e. independently 
of what their neighbors are doing. They do this with probability $\lambda$. In terms of flip rates one has
\[C(x,\rho) = (1 - \lambda)C_v(x,\rho) + \frac{\lambda}{\tau}\left[1 + (1-2p)(2\rho(x) -1)\right],\]
where $0 \leq p\leq 1$ and $ 0 \leq \lambda \leq 1$ and $C_v$ is the
voter model flip rates. This leads to a stationary state in any periodic box of size $L^d$ with  density of pluses
equal to $p$. As $\lambda$ increases from 0 to 1 we go from the voter model to an independent flip model. 
The stationary state of the latter is a product measure with density $p$. This model was studied rigorously in
~\cite{noisy} where it was named the noisy voter model.
\\
\indent
In~\cite{LebowitzSaleur} the authors used (\ref{Scaling}), on simulation results of the noisy voter model
on lattices with periodic boundary conditions, to obtain $p_c(\lambda)$ for $\lambda > 0.1$. 
For $d=3$ they found by extrapolation $p_c(\lambda) \sim 0.16$ as $\lambda \rightarrow 0$. 
\\
\indent
We have repeated the simulations in ~\cite{LebowitzSaleur} for larger lattice sizes and smaller values of $\lambda$. 
We simulated systems with $\lambda$ as small as 0.01 each with
24000 ``effectively uncorrelated'' samples and sizes up to $80$. From our results we can extrapolate 
$p_c(\lambda) \rightarrow 0.15$ as $\lambda \rightarrow 0$, a value
slightly lower than the result in ~\cite{LebowitzSaleur}. We also observed that, as expected, the critical exponents
for the noisy voter model agree, for the given range of $\lambda$, with the critical exponents of independent 
percolation. 
\\
\indent
This leaves a significant difference with the result for $p_c$ obtained
in the previous section. We believe that the answer lies in the necessary extrapolation to $\lambda =0$. Since
the autocorrelation time grows exponentially
with lambda, this means we have to wait for more and more Monte Carlo steps to get independent samples.
To check this explanation we investigated the percolation
transition in the harmonic crystal with a mass M. This mass acts much like the random flips in the voter model. 
For both models the pair correlation decays exponentially. In the harmonic crystal the characteristic
length scale is $\xi_M = \frac{1}{M}$. An easy calculation shows that the characteristic length scale for the
noisy voter model is $\xi_{\lambda} = \sqrt{\frac{1-\lambda}{6 \lambda}}$.
The noisy voter model with the smallest lambda that we simulated , $\lambda = 0.01$,  
thus corresponds to $\xi_{\lambda}$ roughly equal to 4 (unit distance is the lattice spacing). 
In the language of the massive harmonic crystal this corresponds to $M \sim 0.25$. 
Estimating the percolation threshold of the massless harmonic crystal by the extrapolation
method we used for the voter model using $M \geq 0.25$  yields a $p_c(M) \sim 0.21$ when $M \rightarrow 0$. 
This is obviously a large overestimate of $p_c=0.16$ which was obtained
by directly simulating the massless harmonic crystal. This shows that the extrapolation method greatly overestimates
the true $p_c$. 
\section{Concluding Remarks}
\indent
We have performed Monte Carlo simulations to obtain the critical percolation density and some critical exponents
for the massless harmonic crystal and the voter model in $\mathbb{Z}^3$. We found, for the first time a value of 
$p_c$ for the former and using a novel method of simulation for the voter model found a new more reliable value of $p_c$ 
for this system. The critical exponents for both models 
agree within the error bars. This suggests that both percolation models are in the same universality class and confirms the 
theoretical predictions made in ~\cite{WeinribHalperin}. The result for the correlation length critical exponent $\nu = 2$
supports the conjecture by WH that the relation $\nu = \frac{2}{a}$ is exact. 
\\
\indent
It is believed that not only the critical exponents but also the finite size scaling functions are universal. While this
is certainly consistent with our simulations we have not checked this carefully. Such a check would require measuring quantities for the 
two systems in the same way. This is not what we have done here as we wanted to use the most efficient method for each
system.
\\
\indent
We mention here that there has been much activity in generalizing the voter model in various ways ~\cite{voterclass}. 
Based on our  present work we expect that the nature of the percolation transition in these models will depend only on 
the asymptotic behavior of $G(r)$. We have however not investigated this. Our simulation method may be extendable to some
of these systems. 
\\
\indent
The reported simulations were done on a Sun Microsystems HPC-10000 system.
\section*{Acknowledgments}
We thank J. Cardy, G. Giacomin, P. Ferrari and A. Sokal for useful discussions. We also thank CAIP at Rutgers University, New Jersey for providing 
computing resources. This research was supported in part by NSF Grant DMR-044-2066 and AFOSR Grant AF-FA 9550-04-4-22910. 

\newpage
\input{epsf}

\input{epsf}
\input{epsf}

\input{epsf}

\input{epsf}
\input{epsf}
\input{epsf}
\input{epsf}
\end{document}